# A Survey on Domain-Specific Languages for Machine Learning in Big Data

Ivens Portugal
David R. Cheriton School of Computer Science
University of Waterloo
Waterloo, Canada
iportugal@uwaterloo.ca

Paulo Alencar
David R. Cheriton School of Computer Science
University of Waterloo
Waterloo, Canada
palencar@cs.uwaterloo.ca

Donald Cowan
David R. Cheriton School of Computer Science
University of Waterloo
Waterloo, Canada
dcowan@csg.uwaterloo.ca

*Abstract*—The amount of data generated in the modern society is increasing rapidly. New problems and novel approaches of data capture, storage, analysis and visualization are responsible for the emergence of the Big Data research field. Machine Learning algorithms can be used in Big Data to make better and more accurate inferences. However, because of the challenges Big Data imposes, these algorithms need to be adapted and optimized to specific applications. One important decision made by software engineers is the choice of the language that is used in the implementation of these algorithms. Therefore, this literature survey identifies and describes domain-specific languages and frameworks used for Machine Learning in Big Data. By doing this, software engineers can then make more informed choices and beginners have an overview of the main languages used in this domain.

*Keywords—literature survey; domain-specific languages; DSL; Machine Learning; ML; Big Data; BD*

## I. Introduction

The evolution of technology in the last decades caused a major data revolution [72]. Every second, smartphones, tablets, cars, websites, and systems generate a massive amount of data, and users and software engineers have access to a subset of that data to perform their activities. For example, a user who is about to start a trip may inspect a digital map with the visual displacement of roads and cities to decide the best route to take, based on real-time traffic and weather information. In addition, a software engineer may have access to a user's previous trips and routes taken, to develop a system that is capable of recommending the best route based on that information. This recent phenomenon of the generation of a high volume of digital data that is available to be processed is called Big Data [4].

Big Data not only accounts for the large *volume* of data, but also for the unstructured nature of that data, known as *variety*, and the need of immediate processing of this data, known as *velocity*. These three V's of Big Data were introduced by Doug Laney [21] and are generally used to define Big Data and its characteristics. Big Data also created new challenges in data management. Traditional ways of data storage and analysis do not scale well to this amount of data, which can reach hundreds of terabytes or more, and new approaches are being developed to address these issues [36].

Analyzing this data is relevant because it gives insights about the reasons and the context that data is generated [55]. Additionally, it helps analysts estimate future trends about the use or the context of data. For instance, when analyzing data generated by smartphone sensors (accelerometer, gyroscope, heart monitor), a system can discover the exercise patters of a user. Inferences about this data can help suggest better exercise approaches, as well as warn about incorrect exercises.

Machine Learning is one technique of processing data and making inferences about it [59]. This research area is being widely used to discover patterns, identify trends, suggest actions, and optimize output. Because Machine Learning techniques make better inferences and predictions when more relevant data is available, applying these techniques to Big Data field is of major interest [56].

However, applications of Machine Learning in Big Data have to overcome several Big Data challenges [38]. Data storage and processing should be carefully considered, and algorithms may be described in a parallel way [2]. Data may be scattered across different machines, perhaps distant from each other, and computations may be performed locally at each machine, to later be aggregated and to generate an output.

The implementation of Machine Learning techniques in Big Data may have additional challenges. Some approaches that ease this implementation are required. One of these approaches is the use of a Domain-specific Language for the implementation of the aforementioned techniques. Domain-specific Languages are languages whose instructions are easy and intuitive for a specific domain [23]. For example, HTML and SQL are considered domain-specific languages for respectively webpages creation and relation database querying. Languages that are specifically created to account for the development of Machine Learning techniques in Big Data have high-level abstractions that reduce the focus on overcoming low-level challenges.

This work presents a survey on Domain-specific Languages for Machine Learning in Big Data. It identifies and discusses some languages that were developed for this purpose. This is relevant for software engineers that are about to start the implementation of a system or application that uses Machine Learning techniques in massive amounts of data. This work is also especially relevant for beginners in the field, who do not have experience with the languages and need to choose one to learn. In both scenarios, software engineers can make more informed choices based on this overview of the languages.

This document is organized as follows. In Section 2, it is discussed the Big Data research field, its applications and challenges. In Section 3, the Machine Learning research field is defined and described, as well as the importance of its relationship with Big Data. In Section 4, this work explains Domain-specific Languages, gives some examples, and presents the classification used in the survey. Section 5 lists the languages used in the survey and provide a description about them. Section 6 concludes this work with some insights and future work.

## II. BIG DATA

The Big Data research field emerged from the evolution of technology [4]. In the past, the number of digital devices was not significant, when compared to the quantity of smartphones or tablets (and other so called smart devices) that exist in today's world [53]. As a consequence, the amount of digital data produced was relatively low. Web services and scientific computation produced less data and were slower [30][57] decades ago. The scientific breakthroughs of the recent years made possible the generation of a massive amount of data and in a fast pace, that needs to be stored, analyzed and visualized. In Big Data, traditional approaches of data storage, analysis, and visualization do not work well because they do not scale up, and thus new approaches are being created and studied. In an effort to characterize Big Data, Doug Laney introduced a 3V model [21]:

- Volume - relates to the amount of data generated, which can be in the order of hundreds of terabytes.

- Variety - relates to the number of sources and formats of data, which can be text, images, videos, and even streams of data.

- Velocity - relates to the speed of data processing, which can be as fast as real time

As examples, Walmart, the chain of hypermarkets and grocery stores, registers 267 million transactions each day [15] across its 6000 stores worldwide. Additionally, the Large Hadron Collider (LHC), the particle collider built on the border between France and Switzerland to study and test several theories of particle physics, generates 60 terabytes of data each day [9], that needs to be processed for scientific purposes.

Big Data has applications in several domains that range from Geography to Business. The analysis of data is important and is being used to predict trends in society, or estimate and optimize costs and profits. Some examples of works being developed in Big Data research field focus on making better weather prediction [11], helping in traffic management [43], creating smart cities [34], helping in analysis of diseases [58][61] and social networks [33][69], and helping decision making in business intelligence [22][42].

To handle the massive amount of data efficiently, the Big Data research field has several challenges, which relates to data capture, storage, analysis, and visualization [15][29]. Data capture challenges are mainly associated with devices not always being able to produce data, or data in the real world that is difficult to capture. For example, the GPS device on a smartphone may not be always working because the low coverage in the region where the user currently is in. Moreover, an application that monitors the amount of calories consumed by users may find that the calorie intake data of a user is difficult to measure. Overcoming these challenges is of extreme importance, because data capture is the first step to deliver services or to make analysis.

Data storage challenges refers to ways of storing huge amounts of data in databases. The traditional relational databases do not scale well to Big Data and novel approaches are being used. Research in both distributed storage, in which the database is split in several disks or even in several machines, and NoSQL databases, in which some database properties are relaxed for storage and query, are being developed to account for these challenges. Once data is stored, it is analyzed for insights.

Data analysis challenges relates to processing and generating insights from the massive amount of data stored. It can be particularly difficult and inefficient to process hundreds of entries in a database to find specific information requested by the use of a system. For example, a user, typically a data scientist, who wants to find a list of users interested in a particular product, for marketing purposes, may query the system, say a social network, to return users who mentioned a certain brand in their messages in the last year. The system may have to check all the messages sent in the last year by all users of the social network, and look specifically for the string containing the name of the brand to be advertised. To address this challenge, studies are being developed in the areas of parallel, distributed and cloud computing, where computations like the query exemplified before is replicated in several processors, disks or even machines.

Data visualization challenges refers to ways of representing knowledge intuitively and efficiently, using different graphs. Visualizing data is the primary and most important way of having insights about the present and future of some phenomena. However, with the amount of data generated in Big Data, visualization needs to be carefully planned and executed. A well-constructed graph can show where, when, or how (temporal, spatial, or situational condition) data behave. Scientists then are able to draw conclusions and predict trends. Research on these visualization challenges can be found on [54][64][75].

In the near future, two areas are increasingly relevant to Big Data researchers: privacy and Internet of Things. Several applications of Big Data collect and process user data, either to provide services for the user, such as recommendation systems or information search, or analysis, such as testing scientific theories or decision-making in business intelligence. For that reason, several studies analyze the privacy implications [16][40][60] of these approaches, and ethical questions about data ownership. As a result, some limits of data collection may arise. Another future direction [14] shows that Big Data will become even more important with the development of the

Internet of Things (IoT) research area in the next few years. In IoT, there are several systems and sensors connect to deliver better results to users. For example, buses and taxis can have an embedded GPS device generating location data, which can be presented to users on a mobile application. Users then have access to real time location of buses and taxis, and are able to plan a trip to another city more efficiently. In IoT, even more data will be generated and more studies in Big Data needs to be performed to account for the new challenges.

III. MACHINE LEARNING

The Machine Learning research area has roots in the middle of last century, but just became popular in the 1990s. Today, it is one of the most important research fields. Scientific achievements in this area are changing the way humans interact with computers and promising results are to come in the next decades.

A useful, and somewhat formal, definition of Machine Learning made by Tom Mitchel [48] states: "A computer program is said to learn from experience E with respect to some classes of tasks T and performance measure P, if its performance at tasks in T, as measured by P, improves with experience E". This definition clearly states that there is a positive correlation between performance and experience in Machine Learning. This means that Machine Learning applications may increase their performance and accuracy according to the amount of data they are exposed to.

Machine Learning applications are currently present in many domains that go from entertainment to the analysis of businesses. Microsoft's Kinect [44], for example, is a very popular movement recognition device for Microsoft's Xbox [45]. It has Machine Learning algorithms that can identify body parts of users using a camera. The main applications of this technology is in the video game industry, but other research studies use it, for instance, to detect Parkinson's disease [65], improve limb rehabilitation [19], or study balance and risk of falling in older adults [37]. Autonomous driving [20][32] is another application of machine learning techniques that has been under investigation and is generating interesting results. Major car companies and manufactures around the world are heavily researching and investing in this technology. Recommendation systems also use Machine Learning in their workings to better recommend items to users. Several studies in the literature [39][66][74] describe approaches in this direction.

Machine Learning algorithms are classified in supervised, unsupervised, and reinforcement learning [56]. Supervised learning algorithms have a training phase, where labeled data, i.e. data with the expected output, is fed into the algorithm, so it can learn. On a second phase, the algorithm is executed having unlabeled data as input and generating output. Some examples of algorithms are Logistic Regression [24] and Neural Networks [76]. Unsupervised learning algorithms works on unlabeled data only and uncover hidden patters. Some algorithms that can be used in unsupervised learning are Clustering [50] and k-means [28]. Reinforcement learning algorithms receive feedback from the real world at each output, and based on it, they may improve their future outputs. An analogy is learning from the mistakes. Some examples of reinforcement learning algorithms are Bandits [31] and Q-learning [70].

Applying Machine Learning in the Big Data field is desired and beneficial [73]. The assumption is that the more data is available to be processed, the better are the accuracy of algorithms in predictions. However, such algorithms have to overcome Big Data problems of storage and processing. Because data is large in size and amount, they are stored in different locations and its retrieval is not easy. Moreover, processing that data may not be efficient, and algorithms have to be adapted and optimized to specific cases [2][26].

One key element in the development of Machine Learning algorithms in Big Data is the language used. Depending on the choice of the development language, some abstractions may be easy represented and quickly described. For example, a library that helps software engineers execute distributed matrix multiplication speeds up algorithm development, since the software engineer do not need to spend time or effort into overcoming Big Data low-level challenges for this operation. To account for challenges like this one, this document surveys Domain-specific Languages used in the development of Machine Learning algorithms that work on Big Data. More details are explained in the next section.

IV. DOMAIN-SPECIFIC LANGUAGES

One fundamental part of development of systems in computer science is the communication of ideas, either between humans or from humans to machines. To represent these ideas, software engineers use a language. Although most of the languages used by software engineers are programming languages, such as C or Java, the set of useful languages are not limited to only this type. Software engineers may use modeling languages to represent a system, or communicate a concept among them. Modeling languages can be described using figures or drawings made by hand or with the aid of computer software. Moreover, prior to the development of a system, software engineers communicate the requirements of the system, i.e. a description of the needs of the stakeholders that the system to be developed will solve or address. Requirements are often written in natural language, but alternatively they can be written using formal language. Languages in computer science can be separated into general-purpose languages (GPL) or domain-specific languages (DSL). GPLs are languages that are used to solve problems in several domains of computer science. Some examples of GPLs are C, Java, or Python. These languages can be used to create a system to manage academic information in a university, or to control equipment in the space. In contrast, DSLs are languages created to address problems of a specific domain. Their expressiveness is focused into a specific class of applications [23]. Some popular DSLs are SQL, HTML, Lisp, and OpenGL. These languages focus on a subset of the whole problem space in computer science. SQL, for example, is used to retrieve data from a relational database. Its expressiveness is limited to the most useful operations in this domain. This limited expressiveness has a positive point. It moves the software engineer away from low-level details that are not the focus of the system in development, and let them target at the work of the solution for the problem at hand. In reality, there is a fuzzy boundary in the classification between GPLs and DSLs, so an agreement may not always be possible. For example, a so-called GPL that has several libraries to facilitate and automate graph processing may be considered a DSL for the network domain.

Domain-specific languages have several advantages over general-purpose languages [17]. DSLs offer pre-defined abstractions to represent concepts from the application domain. This representation may be more clear and intuitive. Moreover, DSL compilers may optimize the code written for the specific domain, and they can perform error detection more efficiently. Lastly, DSLs may have more specific tool support that help software engineers increase their productivity.

This document shows a survey on DSLs used in the domain of Machine Learning in Big Data. The choice of limiting the target of the survey only to DSLs is to identify some languages that are used in the domain, but are not well known as the GPLs. However, a survey on the GPLs used in this domain is planned as a future work. Additionally, this survey considers three types of languages: programming, modeling, and requirements.

To describe the languages selected to this survey, a research on the classification of DSLs was performed. The features described in three different publications [23][67][68] compose the classification scheme that is used in this survey. The DSLs considered in this survey are described according to the features explained as follows. DSLs may assume the three following types: *requirements*, *programming*, or *modeling*. Requirement languages express what must be present in a system, while programming languages are used to implement the system. Modeling languages are usually graphical languages used to express ideas, algorithms or components of a system, and languages of this type are also considered in this survey. The nature of the language may be *textual* or *graphical*, depending on the way it is expressed. DSLs can be *internal* or *external*. The former one uses a host language to be developed. Most of the times, an external DSL inherits part of the GPLs syntax, and leverage the GPLs' tools such as the compiler, and the running environment. The host language is usually, but not necessarily, a GPL. In contrast, external DSLs are languages created from scratch. The developer of the language has the freedom to specify the syntax, the default library, and the intuitiveness of the language. Because external languages do not have a base language, its developer has to decide the language's *target platform*, which is usually a compiler, and its *execution engine*. DSLs are also classified into *dynamically typed*, when variable type checking is done at runtime, or *statically typed*, when that operation is done in compilation time. According to the way DSLs are structured, they can be *imperative* or *declarative*. Imperative DSLs have instructions that describe how data is manipulated during execution as a sequence of steps. Declarative DSLs, although still a cloudy term, refer to the languages that move away from imperative style and express what should be executed, putting less attention to how it is done. For that reason, DSLs are considered to be in a higher level of abstraction. Finally, domain-specific programming languages can be *translated* or *interpreted* to machine code. Translation, or compilation, is the process of applying syntax and semantic analysis of the code, as well as some optimization routines, prior to generating the final executing code. Interpretation uses pre-compiled routines to quickly and immediately execute each line of code written. Domain-specific modeling languages also have two ways to be classified. *Descriptive* modeling refers to models that express concepts of a system. They abstract some aspects of the system, such as components, calculations, or the relationship with users, and emphasize others. Alternatively, a *prescriptive* modeling relates to models that can be used to automatically construct the target system. This means that a prescriptive modeling language may be written in a level of formality that enables the software engineer to generate programming code that will likely be used in the development of the system.

V. THE SURVEY

This paper presents a survey on Domain-specific Languages for Machine Learning in Big Data. DSLs ease the implementation of Machine Learning algorithms with the use of high-level abstractions or reusable pieces of code that hides low-level details from software engineers, letting them focus on the main problem at hand. By analyzing languages that are focused in this domain, software engineers can make more informed choices before starting an implementation of an algorithm, and beginners may learn what are the most used and main languages of the domain. Table I, at the end of this paper, lists the DSLs that were identified and summarizes the classification of the languages using the properties explained in the previous section. Each DSL of the list is described in the following paragraphs.

OptiML [12][25][62] is a DSL designed as a research project from Stanford University's Pervasive Parallelism Laboratory with the goal of enabling Machine Learning algorithms to take advantage of parallelism. Its authors expect to bridge the gap between Machine Learning and heterogeneous hardware of Big Data without compromising the productivity of software engineers, neither the performance of the algorithm. OptiML is a textual programming language built on top of Scala [49], another DSL that is explained in more detail later. Variables in OptiML have their types specified before execution, which means that the language is statically typed. Moreover, OptiML is declarative because its high-level code constructions hide low-level parallelization concerns, which makes the code clear and efficient. OptiML runs on top of Delite [63], a compiler framework developed by the same research team, and supports operations with the basic three types: vector, matrix, and graph. Its operations supports parallel executions (using the MapReduce programming model [18]) in heterogeneous machines, which are machines that have more than one type of processor or core. One limitation of the language is its lack of support for a distributed environment or executions in the cloud.

ScalOps [8][71] is a DSL with the goal of enabling Machine Learning algorithms to run on a cloud computing environment and overcoming a limitation of the traditional MapReduce programming model: the lack of iteration. ScalOps is a textual programming DSL developed in jointly by the University of California, Irvine and Santa Cruz, and the division Yahoo! Research. The DSL also has Scala [49] language serving as a host language, which means that ScalOps is an internal DSL. Its high-level syntax makes it a declarative language, and as the type checking happens in compilation-time, ScalOps is considered a statically typed. ScalOps needs to be compiled to generate lower level code, which makes it be classified as a translated language, according to the analysis of this survey. Additionally, the language supports vector, matrix, and graph operations in both parallel and cloud computing environment. To support iterations in MapReduce, ScalOps designers introduced an enhanced version of the programming model called Map-Reduce-Update [8]. This new version consists of three user-defined functions called map, reduce, and update. The map function receives read-only global state values and is applied to training data points in parallel. The reduce function aggregates the output of the map function. Finally, the update function receives the aggregated value and produces a new global state value for the next iteration. Alternatively, when appropriate, the update function indicates that no additional iteration is necessary.

Pig Latin [51] was also developed by Yahoo! Research. Released in 2008, the language has the goal of abstracting MapReduce implementation and turning it easy to program and to perform. Pig Latin runs on top of Apache Pig, a platform developed by Apache and it implements Apache Hadoop [3]. Hadoop is the framework used by Pig to execute MapReduce. Pig Latin is a textual programming DSL that has no host language, which means that it is classified as an external DSL. The language is dynamically typed, and variables can assume the form of an atom, for atomic values, tuple, for a sequence of fields, bag, for a collection of tuples, or map, for a collection of data items where each item has an associated key. Although Pig Latin follows the declarative style of SQL, it is considered a procedural language because of the way software engineers describe operations on data. Pig Latin features a compiler that translates the DSL code into MapReduce jobs, which are then executed using the existing Apache Hadoop infrastructure under Apache Pig platform. Pig Latin offers built-in support for vector and matrix operations. It also assists parallel execution of algorithms. Machine Learning algorithms can be implemented using user-defined functions, which can be written in the following languages: Java, Python, Javascript, Ruby, or Groovy. Pig Latin also supports distributed and cloud computing by leveraging from Apache Hadoop Cluster [3], a framework for distributed storage and processing of very large data sets.

SCOPE [13] is a DSL developed by Microsoft to account for the distributed and cloud computing challenges in querying and managing data. SCOPE stands for Structured Computations Optimized for Parallel Execution, and was also released in 2008. The language simplifies parallel processing of massive data sets without compromising efficiency. It is a textual programming language and an external DSL, with its own syntax and no host language. Regarding SCOPE's syntax, it is similar to SQL with very similar instructions such as *select*, *from*, *where*, *group by*, and *order by*. SCOPE is dynamically typed, i.e. its variables have types assigned during execution time. Custom user-defined functions are written in C#, which is a GPL and is not considered in this survey of the literature. SCOPE is a declarative DSL, which means that it have abstractions to allow software engineers to focus on data transformations that are required to solve the problems at hand and it also mean that the syntax of the language hide the complexity of the underlying platform and some implementation details. SCOPE code is compiled by a SCOPE compiler and executed by Cosmos Execution Environment. Data is stored in Cosmos Storage System [13], also developed by Microsoft. Parallel computations are supported by SCOPE, and they are executed using a MapReduce-like programming model. The new model uses two native functions: process, which takes row sets as inputs, processes it, and outputs another sequence of rows; and reduce, which groups data columns, process them, and outputs zero, one or multiple rows per group. Only vector and matrix operations are natively supported by SCOPE, but computations may happen in parallel, distributed, or cloud environments.

Sawzall [52] is a DSL developed by engineers at Google. It addresses the problem of retrieving and processing data separated across several disks in a MapReduce fashion, with filters and aggregators. Sawzall is a textual programming DSL first described in 2003. The language is an external DSL, having no host language, and statically typed. According to the authors, the reason for creating the language with a static type model is to avoid rework in case of type error. Executing scripts in Sawzall, and in Big Data in general, can be very time-consuming and the cost of a late-arising dynamically typed error is very expensive. Sawzall is also classified as an imperative language, which means that code written should describe how data is manipulated, and not what is done to manipulate the it. Sawzall code is interpreted into low-level language, and only the Sawzall runtime, called szl, was open sourced. The Sawzall engine remains proprietary. Data manipulated in Sawzall are stored in the Google File System (GFS), a proprietary distributed file system created by Google to store the large amount of data the company deals with. Finally, Sawzall language offers parallel, distributed, and even cloud computations, and supports the basic types vector and matrix.

The next language in the list did not have a name associated with it, and it was decided to refer to it as VisuML [6]. VisuML is a language aimed at representing, in high-level, Machine Learning systems that handle Big Data. The language is based on abstractions of other modeling languages and some language constructs, namely graphical models [27], factor graphs [5], plates[10], and gates [46]. The author of VisuML explains the syntax of the language but do not offer a translation to any programming language. Instead, the author only suggests a way of implementing VisuML's constructions in C#. The actual implementation is left as future work. Therefore, VisuML is classified as a modeling DSL. It uses its own symbols to express domain-specific concepts, which makes it an external DSL. Finally, because its symbols cannot generate code, VisuML is said to have a descriptive model.

Graphical models [27] are a language to describe probabilistic models visually. Circles represent random variables and the relationship between them, which is expressed with directed line segments, represents joint distributions. The language is widely used to describe Bayesian networks, which is a probabilistic model used to perform inference and learning in data. Graphical models are classified as a graphical and modeling DSL that do not have a host language, which means it is external. Additionally, the language has a descriptive model, since its graphical representation is not meant to automatically generate code.

This survey's goal is to identify DSLs that address problems in the domain of Machine Learning in Big Data. However, another important component of the solutions for these problems is the framework that works in the execution of the language. It was decided to include in this survey a brief description of some frameworks found in the literature, as it can help in the decision of the language used to express or develop systems in the domain addressed by this work. Table II, also shown at the end of this document, lists the frameworks alongside its characteristics. A more detailed explanation is given in the following paragraphs.

Infer.net [47] is a framework created by Microsoft Research Cambridge aimed at running Bayesian inference in graphical models. Users write their textual codes in any .NET language, such as C#, C++, F#, or IronPython, and then they run their code to make inferences on data. The framework offers built-in support for vector and matrix operations, but lacks to assist users in parallel, distributed or even cloud computing.

Graphlab [41] is a framework for the implementation of Machine Learning algorithms, developed jointly by researchers in Carnegie Mellon University and the University of California Berkeley. Its main focus is to provide parallelization of Machine Learning algorithms and thus increase the efficiency of their executions. Algorithms are written using textual code in C++ or Python. Although the framework originally offered assistance to only graph operations, its creators later added support to vector and matrix operations. The framework also supports code execution in parallel, distributed, and cloud computing environments. Years after GraphLab framework project was started, in 2009, it was managed by the startup GraphLab.Inc. Today, the startup evolved to Dato.Inc, and the framework has a commercial version called GraphLab Create.

TensorFlow [1] is a framework introduced by Google Research in 2015 to express and execute Machine Learning algorithms efficiently in scalable manner. The need for efficiency and scalability is to address the Big Data challenges explained in previous sections. Internal services at Google use TensorFlow, and according to the authors the execution of algorithms process data containing hundreds of billions of parameters and in hundreds of heterogeneous machines. Machine Learning algorithms in TensorFlow are written in either C++ or Python. The framework supports users with vector, matrix, and graph operations for parallel execution. Distributed and cloud computing are also assisted by TensorFlow.

## VI. CONCLUSIONS

In the last decade, the number of digital data generated by devices, services, and scientific applications increased significantly and new problems associated with this phenomenon arose. Traditional methods for data capture, storage, analysis, and visualization did not scale well to the new reality, and research on novel methods is being performed to address the new problems. These problems, their discussion, and the solutions proposed in the literature created a new research area called Big Data. Big Data is mainly characterized by the presence of a large amounts data (volume), but also by the great number of sources and formats (variety), and by the speed of processing required (velocity). Data have important properties and relationships that are key to new research opportunities and for business strategy. Therefore, analyzing and inferring these properties is paramount researchers and practitioners for the near future.

Machine Learning is a research area that is reshaping the way humans leverage from computational analysis. It describes algorithms that not only analyzes data, but also predict trends, and recommend decisions to be taken. These algorithms are mathematically efficient, but have to deal with Big Data problems when implemented. Data size is large and may be stored in several disks, which belong to several machines, geographically distant. Additionally, computations may have to be distributed in this infrastructure especially for performance issues. Therefore, in most of the times, code needs be optimized to each application being developed, so efficient methods of data retrieval and processing are used by the algorithms. When adapting the Machine Learning algorithms, one important choice that software engineers need to make is the language that will be used. This survey identifies and characterizes DSLs for Machine Learning in Big Data, so software engineers can make better and more informed choices, and beginners can be introduced to some of the languages being used.

This survey identified seven languages published in the literature or being used in the domain of Machine Learning in Big Data, and described them using a classification created from publications about DSL found in the literature. Most of the identified languages are programming DSLs, and no DSL was found target to expressing requirements for systems in this domain. Half of the six programing languages of this survey were indirectly built on top of Java, and the remaining ones define its own syntax, i.e. they are external DSLs. The number of statically and dynamically typed languages is also balanced, and so is the number of declarative or imperative DSLs. However, the great majority of them are translated (compiled) when generating low level code, while only one is interpreted. All programming languages offers parallel execution of the code that was written, but not all supports distributed or cloud computing. Domain-specific modeling languages considered in this survey are graphical and have a descriptive model, which means that they only express the system, but do not automatically generate its code.

As an additional contribution, this survey also identified and described some frameworks used in the domain of Machine Learning in Big Data. Software engineers can program algorithms in several languages, but based on the languages used by the frameworks researched in this survey, one can conclude that there is a trend of using C++ and Python for the implementation of Machine Learning algorithms in the domain surveyed. Support for vector, matrix, and graph operations in a parallel environment is also observed as a trend, especially among the most recent frameworks described.

Finally, recent frameworks are assisting users with distributed and cloud computing.

This survey is not extensive, and more languages and frameworks can be described and added in the near future. One future work in this direction is to survey general-purpose languages used in the domain of Machine Learning in Big Data. These languages, although not specifically created for the domain, have may have some properties that are critical to some types of applications, such as real time Machine Learning in Big Data. The identification of the languages and frameworks in this survey represents an important step in the knowledge of the domain, which can improve and revolutionize the way society interact with computers in the future.

VII. ACKNOWLEDGMENTS

The authors would like to thank the Natural Sciences and Engineering Research Council of Canada (NSERC), the Ontario Research Fund of the Ontario Ministry of Research and Innovation, SAP, and the Centre for Community Mapping (COMAP) for their financial support to this research.

TABLE I. DSLs FOR MACHINE LEARNING IN BIG DATA

| Language Name | Requirements / Programming / Modeling | Textual / Graphical | Internal / External | Dynamically typed / Statically typed | Imperative / Declarative | Translation / Interpretation | Target Platform / Execution Engine | Descriptive model / Prescriptive model | Supports Vector (V) / Matrix (M) / Graph (G) operations | Supports Parallel operations | Supports Distributed (D) / Cloud (C) computing |
|---|---|---|---|---|---|---|---|---|---|---|---|
| OptiML [62] | Programming | Textual | Internal (Scala) | Statically typed | Declarative | Translation | - | - | V/M/G | Yes | -/- |
| ScalOps [71] | Programming | Textual | Internal (Scala) | Statically typed | Declarative | Translation | - | - | V/M/G | Yes | D/C |
| Pig Latin [51] | Programming | Textual | External | Dynamically typed | Imperative | Translation | Pig Latin compiler / Apache Pig | - | V/M/- | Yes | D/C |
| SCOPE [13] | Programming | Textual | External | Dynamically typed | Declarative | Translation | SCOPE Compiler / Cosmos Execution Environment | - | V/M/- | Yes | D/C |
| Sawzall [52] | Programming | Textual | External | Statically typed | Imperative | Interpretation | Sawzall compiler / Sawzall engine (proprietary) | - | V/M/- | Yes | D/C |
| VisuML [6] | Modeling | Graphical | External | - | - | - | - | Descriptive | - | - | - |
| Graphical models [27] | Modeling | Graphical | External | - | - | - | - | Descriptive | - | - | - |

TABLE II. FRAMEWORKS FOR MACHINE LEARNING IN BIG DATA

| Framework name | Textual / Graphical | Languages | Supports Vector (V) / Matrix (M) / Graph (G) operations | Supports Parallel operations | Supports Distributed (D) / Cloud (C) computing |
|---|---|---|---|---|---|
| Infer.net [47] | Textual | .NET framework languages | V/M/- | Yes | -/- |
| Graphlab [41] | Textual | C++, Python | V/M/G | Yes | D/C |
| TensorFlow [1] | Textual | C++, Python | V/M/G | Yes | D/C |